\documentclass[reprint, aps, pra, superscriptaddress]{revtex4-1}

\usepackage[latin1]{inputenc}
\usepackage[T1]{fontenc}
\usepackage{gensymb} 
\usepackage[seperr,load={}]{siunitx}
\usepackage{lmodern}
\usepackage{amsmath}
\usepackage{amssymb}
\usepackage{graphicx}
\usepackage{color}
\usepackage{soul}

\begin{document}

\title{Hole weak anti-localization in a strained-Ge surface quantum well} 
\author{R. Mizokuchi}
\author{P. Torresani}
\author{R. Maurand}
\affiliation{CEA, INAC-PHELIQS, F-38000 Grenoble, France}
\affiliation{University Grenoble Alpes, F-38000 Grenoble, France}
\author{M. Myronov}
\affiliation{Department of Physics, University of Warwick, Coventry CV4 7AL, United Kingdom}
\author{S. De Franceschi}
\affiliation{CEA, INAC-PHELIQS, F-38000 Grenoble, France}
\affiliation{University Grenoble Alpes, F-38000 Grenoble, France}

\begin{abstract}
We report a magneto-transport study of a two-dimensional hole gas confined to a strained Ge quantum well grown on a relaxed Si$_{0.2}$Ge$_{0.8}$ virtual substrate. The conductivity of the hole gas measured as a function of a perpendicular magnetic field exhibits a zero-field peak resulting from weak anti-localization. The peak develops and becomes stronger upon increasing the hole density by means of a top gate electrode. This behavior is consistent with a Rashba-type spin-orbit coupling whose strength is proportional to the perpendicular electric field, and hence to the carrier density. By fitting the weak anti-localization peak to a model including a dominant cubic spin-orbit coupling, we extract the characteristic transport time scales and a spin splitting energy of $\sim$1 meV. Finally, we observe a weak anti-localization peak also for magnetic fields parallel to the quantum well and attribute this finding to a combined effect of surface roughness, Zeeman splitting, and virtual occupation of higher-energy hole subbands.
\end{abstract}

\maketitle

Hole spins in p-type SiGe-based heterostructures are promising candidates for quantum spintronic applications \cite{RevModPhys.85.961,PSSA:PSSA201600713}. They are expected to display a relatively small in-plane effective mass \cite{sawano2009strain,doi:10.1063/1.4962432}, favoring lateral confinement, as well as long spin coherence times \cite{doi:10.1021/nl501242b}, stemming from a reduced hyperfine coupling (natural Ge is predominantly constituted of isotopes with zero nuclear spin and holes are less coupled to nuclear spins due to the p-wave symmetry of their Bloch states \cite{testelin2009hole}). In addition, low-dimensional, SiGe-based structures benefit from a strong and electrically tunable spin orbit coupling \cite{doi:10.1063/1.4962432,doi:10.1021/acs.nanolett.6b02715,PhysRevLett.110.046602,PhysRevB.84.195314,doi:10.1063/1.4901107,zarassi2016magnetic}. This property could be exploited for purely electrical spin control \cite{doi:10.1021/nl501242b,doi:10.1063/1.4858959,doi:10.1063/1.4963715,maurand2016cmos}. Finally, hybrid superconductor-semiconductor devices based on strained Ge quantum wells confining holes should provide a favorable platform for the development of topologically protected qubits based on Majorana fermions of parafermions \cite{PhysRevB.90.195421,su2016high}.

\begin{figure}[ht]

\begin{center}

\includegraphics[width=1\linewidth]{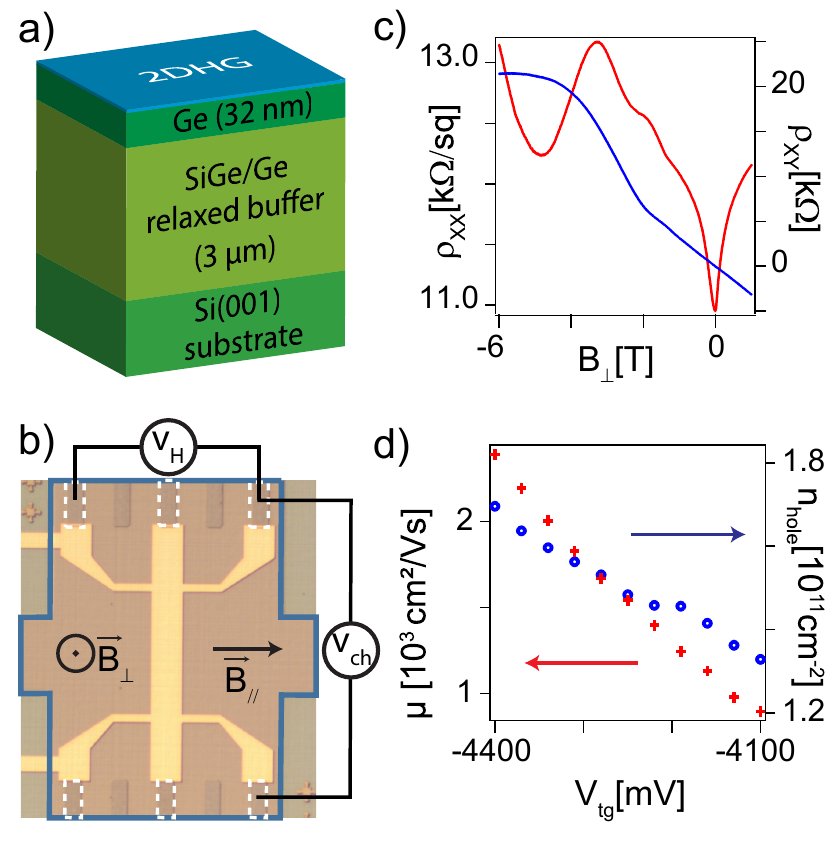}

\end{center}

\caption{\label{fig:hetero} a) Schematic of the heterostructure. The 2DHG (blue) lays on top of the 32 nm Ge layer. b) Optical image of the Hall bar devices. The blue line highlights the mesa and the white dotted lines the Pt contacts. We measure the transverse Hall voltage ($\rm{V_{H}}$) and the longitudinal channel voltage ($\rm{V_{ch}}$) from which we extract Hall resistivity and channel resistivity respectively. The directions of the applied fields $B_{\bot}$ and $B_{//}$ are also indicated. c) Channel resistivity $\rho_{XX}$ (red) and Hall resistivity $\rho_{XY}$ (blue) as a function of out of plane magnetic field. Channel resistivity shows a dip at low field which is a signature of weak anti-localization. d) Mobility $\mu$ (red) and carrier density $n_{hole}$ (blue) as a function of accumulation gate voltage $V_{tg}$.}

\end{figure}

Here we consider a SiGe-based heterostructure with a compressively strained Ge quantum quantum well (QW) at its surface. This heterostructure presents two main advantages: 1) in a metal-oxide-semiconductor field-effect transistor (MOSFET) device layout, it allows for an efficient gating of the accumulated two-dimensional hole gas (2DHG); 2) the surface position of the QW enables an easier fabrication of low-resistive contacts to the 2DHG. The strained SiGe heterostructure was grown on a 200 mm Si(001) substrate by means of reduced pressure chemical vapor deposition (RP-CVD). Growth was realized using an industrial-type, mass-production system (ASM Epsilon 2000 RP-CVD), which is a horizontal, cold-wall, single wafer, load-lock reactor with a lamp-heated graphite susceptor in a quartz tube. RP-CVD offers the major advantage of unprecedented wafer scalability and is nowadays routinely used by leading companies in the semiconductor industry to grow epitaxial layers on Si wafers of up to 300 mm diameter. The heterostructures, shown schematically in Fig. \ref{fig:hetero}.a, consists of a 3 $\rm{\micro\meter}$ thick reverse linearly graded, fully relaxed $\rm{Si_{0.2}Ge_{0.8}/Ge/Si(001)}$ virtual substrate with a 32-nm-thick strained Ge QW surface layer. This is a typical design for surface channel structures employed in modern MOSFET devices. The full structure was grown in a single process without any external treatment. The surface of the Si wafers was cleaned by an in situ thermal bake in $\rm{H_2}$ ambient at high temperature, above 1000\degree C. The Ge epilayer was grown from a commercially available and widely used germane ($\rm{GeH_4}$) gas precursor at a relatively low substrate temperature (<450\degree C), as it is known that the growth temperature of the compressively strained Ge epilayers has to be sufficiently low to suppress surface roughening and retain compressive strain in the epilayers. Further details of materials growth and characterization are described elsewhere \cite{myronov2010high}. The same epitaxial growth technology resulted in the creation of strained Ge QW heterostructures with superior low- and room-temperature electronic properties \cite{doi:10.1063/1.4763476,1347-4065-53-4S-04EH02,6874628} enabling the observation of various quantum phenomena including the fractional quantum Hall effect \cite{PhysRevB.91.241303}, mesoscopic effects due to spin-orbit interaction\cite{morrison2014observation,failla2015narrow,Morrison201684,PSSA:PSSA201600713,0953-8984-27-2-022201} and terahertz quantum Hall effect \cite{1367-2630-18-11-113036}.

The studied devices have a Hall-bar geometry defined by a top-gate electrode operated in accumulation mode (Fig. \ref{fig:hetero}.b). Due to the absence of intentional doping, the Ge QW contains no carrier at low temperature. A sufficiently negative voltage applied to the top gate induces the accumulation of a 2DHG in the QW region underneath. Device fabrication involves the following steps : a relatively large (tens of microns wide), 55-nm-thick mesa structure is initially defined by optical lithography and dry etching in $\rm{Cl_2}$ plasma ; ohmic contacts are successively fabricated using optical lithography, followed by Ar etching (to remove the residual oxide) and Pt deposition in an e-beam evaporator system; 30 nm of $\rm{Al_2O_3}$ are deposited everywhere using atomic layer deposition at 250\degree C ; finally, the Hall-bar-shaped top gate accumulation electrode is defined by e-beam lithography and deposition of 60 nm of Ti/Au. 

Magneto-transport measurements were performed in a $\rm{^3He}$ cryostat with a base temperature of 300 mK. Longitudinal ($\rho_{XX}$)  and Hall ($\rho_{XY}$) resistivities were measured as a function of magnetic field, $B_\perp$, perpendicular to the 2DHG, and top gate voltage ($V_{tg}$), controlling the carrier density. Examples of such traces are given in Fig. \ref{fig:hetero}.c. From Hall resistivity we extracted the hole mobility ($\mu$) and carrier density ($n_{hole}$) ranging from 800 to $\rm{2400}\ {cm^2/Vs}$ and from 1.3 to 1.7 $\times$\SI{e11}{cm^{-2}} respectively, for top gate voltages verying between -4.1 and -4.4V (the threshold $V_{tg}$ for the onset of conduction is at -3.8V, as shown in Fig. \ref{fig:hetero}.d). The mobility is much lower than the one reported in other strained Ge heterostructures \cite{dobbie2012ultra}. This difference is likely due to the presence of charge traps at the Ge/$\rm{Al_2O_3}$ interface.

\begin{figure}[ht]
\begin{center}
\includegraphics[width=1\linewidth]{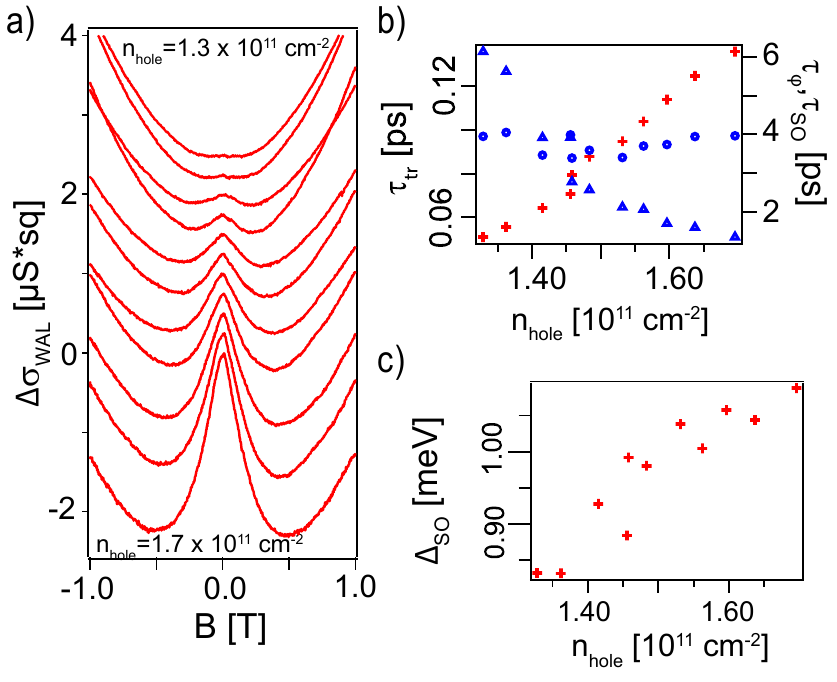}
\end{center}
\caption{\label{fig:magnetoR} a) Traces of the weak anti-localization contribution to the channel conductivity $\Delta\sigma_{WAL}$ as a function of $B_{\bot}$ for different accumulation gate voltages and carrier densities from $1.3\times\SI{e11}{cm^{-2}}$ (top trace) to $1.7\times\SI{e11}{cm^{-2}}$ (bottom trace, traces are offset for better visibility). The weak anti-localization peaks emerges as carrier density is increased. b) Evolution of scattering time $\tau_{tr}$ (red crosses), phase relaxation time $\tau_{\varphi}$ (blue circles) and spin relaxation time $\tau_{SO}$ (blue triangles) as a function of carrier density. c) Evolution of the spin splitting energy $\Delta_{so}$ as a function of carrier density.}
\end{figure}

Following basic Hall-effect characterization we now turn to a more in-depth investigation of the magneto-transport properties. In Fig. \ref{fig:hetero}.c, the longitudinal resistivity (red trace) exhibits a pronounced dip at zero magnetic field. Such a dip is a characteristic signature of weak anti-localization, a mesoscopic phenomenon associated with spin-orbit coupling \cite{PhysRevB.53.3912}. At zero magnetic field the latter leads to a reduced backscattering resulting in a resistivity minimum. This quantum interference effect is suppressed by a magnetic field perpendicular to the 2DHG, accounting for the observed resistivity dip at $B_\perp = 0$. 

This phenomenon is further investigated in Fig. \ref{fig:magnetoR}.a, where the longitudinal conductivity is now plotted as a function of $B_\perp$ and for a range of $V_{tg}$ values, after having removed the feature-less back-ground contribution from classical  Dr\"{u}de conductivity. As a matter of fact, $\Delta\sigma_{WAL}$ represents the quantum correction resulting from weak anti-localization. Interestingly, this data set shows that the weak anti-localization peak develops upon increasing the gate voltage and, correspondingly, the hole density $n_{hole}$ and the perpendicular electric field in the QW. This trend suggests that the spin-orbit coupling responsible for the weak anti-localization effect is of a Rashba type. In our strained QW system, where holes occupy the first heavy-hole subband, we expect the dominant Rashba component to be cubic in the in-plane momentum. \cite{morrison2014observation,moriya2014cubic,failla2015narrow,0268-1242-23-11-114017}. 
 
The weak anti-localization peak can be fitted to the formula \cite{iordanskii1994weak}:




\begin{multline}
\Delta\sigma_{WAL}(B_{\bot})=\frac{e^2}{2\pi^2\hbar}\{\Psi(\frac{1}{2}+\frac{B_{\varphi}}{B_{\bot}}+\frac{B_{SO}}{B_{\bot}})+\frac{1}{2}\Psi(\frac{1}{2}\\+\frac{B_{\varphi}}{B_{\bot}}+2\frac{B_{SO}}{B_{\bot}})-\frac{1}{2}\Psi(\frac{1}{2}+\frac{B_{\varphi}}{B_{\bot}})-\ln{(\frac{B_{\varphi}}{B_{\bot}}+\frac{B_{SO}}{B_{\bot}})}\\-\frac{1}{2}\ln{(\frac{B_{\varphi}}{B_{\bot}}+\frac{2B_{SO}}{B_{\bot}})}+\frac{1}{2}\ln{(\frac{B_{\varphi}}{B_{\bot}})}\}
\end{multline}

where $\Psi(X)$ is the digamma function, $B_{\varphi}$ is the phase coherence field and $B_{SO}$ is the characteristic field associated with the Rashba spin orbit coupling.

From the fitting parameters $B_{\varphi}$ and $B_{SO}$ we can extract the phase coherence time $\tau_{\varphi}$ and the spin relaxation time $\tau_{so}$ with $\tau_i = m^*/4\pi\hbar\mu n_{hole}B_i$, i being either $\varphi$ or SO and $m^*$ being the effective in-plane hole mass (from earlier studies in similar heterostructures \cite{failla2015narrow,morrison2014observation,PhysRevB.89.125401,doi:10.1063/1.4953399}, we assumed $m^*=0.1\ m_0$ where $m_0$ is the bare electron mass). We note that the large width of the observed weak anti-localization peak is consistent with the relatively small values obtained for the scattering time ($\tau_{tr}=m^*\mu/e$). 

These values, as well as those for $\tau_{\varphi}$, $\tau_{SO}$ are displayed as a function of carrier density in Fig. \ref{fig:magnetoR}.b. The evolution of these characteristic time scales with respect to $n_{hole}$ provides a hint  on the underlying mechanism for spin relaxation. If spin relaxation were due to impurity scattering (Elliot-Yafet mechanism \cite{elliott1954theory,yafet1983conduction}), then $\tau_{SO}$  should increase with $\tau_{tr}$ and decrease with the carrier density ( $\tau_{so}\propto\tau_{tr}/n_{hole}^2$). This does not correspond to the observed trend. On the other hand, if spin relaxation occurred in between scattering events, due to spin-orbit-induced rotation (Dyakonov-Perel mechanism \cite{dyakonov1972spin}), the spin relaxation time should decrease with $\tau_{tr}$ and with the spin splitting energy $\Delta_{SO}$ ($\tau_{so}\propto1/(\tau_{tr}\times\Delta_{SO}^2)$, where $\Delta_{SO} \approx \alpha_3 E_z k_F^3$ where $\alpha_3$ is the cubic Rashba coupling, $E_z$ the vertical electric field and $k_F$ the Fermi wave number). Our experimental finding is consistent with this second scenario, which allows us to deduce the the spin splitting energy and its dependence on the carrier density (see Fig. \ref{fig:magnetoR}.c).

The obtained values of the spin splitting energy are around 1 meV, i.e. comparable to those obtained in similar heterostructures and using different experimental methods \cite{morrison2014observation,moriya2014cubic,failla2015narrow}.

\begin{figure}[ht]
\begin{center}
\includegraphics[width=1\linewidth]{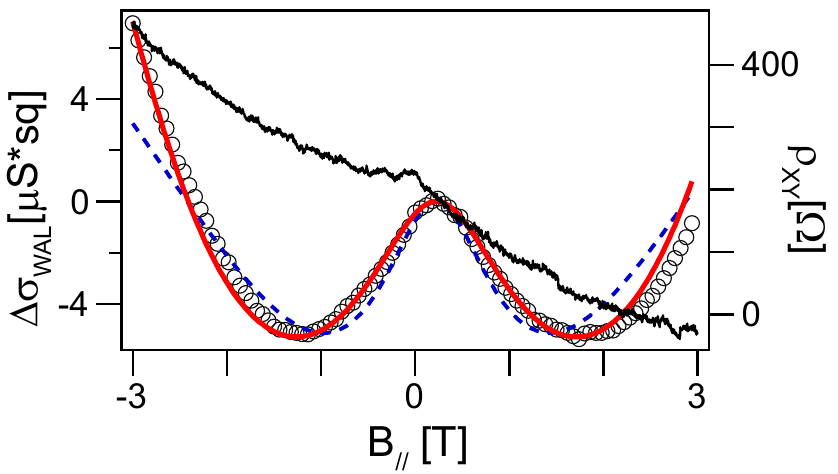}
\end{center}
\caption{\label{fig:inplaneB} Black trace: Hall resistivity $\rho_{XY}$ as a function of in plane magnetic field. The small dependence on field results from a small perpendicular field component. We estimate an angle of only 2\degree\ between $B_{//}$ and the chip plane. Black circles: quantum correction to channel conductivity $\Delta\sigma_{WAL}$ revealing a weak anti-localization peak. The blue dashed line and the red solid line are fits to the model from Minkov \textsl{et al.} \cite{minkov2004weak} without and with the addition of a $B_{//}^6$ term respectively.}
\end{figure}

To further investigate the nature of the zero-field conductivity enhancement, magneto-transport measurements were performed also with the magnetic field applied in the plane of the 2DHG, as indicated in Fig. \ref{fig:inplaneB}. To first order, an in-plane magnetic field is not expected to break the effect of weak anti-localization, because it produces no flux through the time-reversed back-scattering trajectories. Contrary to this expectation, the longitudinal conductivity measured as a function of the in-plane magnetic field, $B_{//}$, does exhibit a clear zero-field peak. The characteristic half width at half maximum of the peak is $\sim 0.7$ T, i.e. several times larger than in the case of perpendicular field. 

We can rule out the possibility of a misalignment of the magnetic field with respect to the plane of the 2DHG. In fact, from a simultaneous measurement of the Hall resistivity, also shown in Fig. \ref{fig:inplaneB}, we estimate a misalignment of only 2\degree. Therefore, the out-of-plane component of the applied field is far too small to explain the observed conductivity peak.

Instead, the effect can be ascribed to the finite thickness of the 2DHG. This type of problem was theoretically investigated by Minkov \textsl{et al.}\cite{minkov2004weak}. According to this work, the weak anti-localization correction to the magneto-conductivity can be expressed as:

\begin{multline}
\label{eq:inplane}
\Delta\sigma_{WAL}(B_{//})=\frac{e^2}{4\pi^2\hbar}\left[ 2\ln{\left(\frac{B_{\varphi}+B_{SO}+\Delta_r}{B_{\varphi}+B_{SO}}\right)}\right.\\+\ln{\left(\frac{B_{\varphi}+2B_{SO}+\Delta_r}{B_{\varphi}+2B_{SO}}\right)}-\ln{\left(\frac{B_{\varphi}+\Delta_r+\Delta_{s}}{B_{\varphi}}\right)}\\ \left.+S\left(\frac{B_\varphi+\Delta_r}{B_{SO}}\right)-S\left(\frac{B_\varphi}{B_{SO}}\right)\right] 
\end{multline}

where $\Delta_r$ and $\Delta_{s}$ are $B_{//}$-dependent  corrections to $B_\varphi$ taking into account the effect of surface roughness and Zeeman splitting, respectively. Following Ref. \cite{minkov2004weak}, we assume $\Delta_r = r B_{//}^2$ and $\Delta_{s} = s B_{//}^2$. The $S(x)$ function in Eq. 2 can be explicitly written as:

\begin{equation}
S(x)=\frac{8}{\sqrt{7+16x}}\left[\arctan{\left(\frac{\sqrt{7+16x}}{1-2x}\right)}-\pi\Theta(1-2x)\right]
\end{equation}

where $\Theta$ is the Heaviside step function. For the effective fields $B_{SO}$ and $B_\varphi$ we take the values extracted from the previously discussed magneto-transport measurements in perpendicular magnetic field, for the same carrier density, i.e. $B_{SO}$ = 170 mT and $B_{\varphi}$= 19 mT. 

The dotted blue line in Fig. \ref{fig:inplaneB} is a fit to equation \ref{eq:inplane} using the proportionality factors $r$ and $s$ as fitting parameters. The fit shows only moderate agreement with the data. A significantly improved fit can be obtained by introducing in the expression of $\Delta_r$ a second orbital term proportional to $B_{//}^6$, i.e. $\Delta_r= r \times B_{//}^2 + q \times B_{//}^6$, with the additional fitting parameter $q$. This second term describes the time reversal symmetry breaking due to magnetic field via the virtual occupation of unoccupied higher energy subbands\cite{0953-8984-2-16-009, PhysRevLett.89.206601,PhysRevLett.89.276803}. The new fit, shown by a solid red line in Fig. \ref{fig:inplaneB}, is in remarkably good agreement with the experimental data set over the entire magnetic-field range. Following Ref. \cite{ PhysRevLett.89.206601}, the value of the fit parameter $q$ can be related to the effective thickness $d$ of the 2DHG, i.e. $ d\sim \left( \frac{q\times \Phi_0^5}{4\pi^2 n_{hole}^2}\right) ^{1/14}$. We find  $d \sim$ 14 nm. Taking into account the thickness of the Ge QW (32 nm) and the gate-induced band bending, this is a perfectly realistic estimate and a validity check for the adopted model.





In conclusion, we have reported magneto-transport measurements of a 2DHG confined to a compressively strained Ge QW on the surface of a relaxed $\rm{Si_{0.2}Ge_{0.8}}$ virtual substrate. The 2DHG is formed by gate-induced hole accumulation up to carrier densities of the order of $10^{11}\ \rm{cm^{-2}}$.

We find that the hole mobility is highly reduced as compared to similar heterostructures where the QW is buried well below the surface. This can be explained by a high density of traps at the Ge/$\rm{Al_2O_3}$ interface, which is in line with the fact that germanium native oxide is known to be of poor quality. In order to enhance hole mobility, a higher interface quality would be required. There exist possible solutions to tackle this problem \cite{zhang2012high,zhang2012high}, which could be explored in forthcoming studies.

Despite the relatively low mobility, weak anti-localization is observed, exhibiting a strong dependence on the carrier density, which is directly related to the perpendicular electric field. This points to a spin-orbit coupling of the Rashba type, with an expected dominant cubic component. The estimated characteristic times $\tau_{tr}$, $\tau_{so}$ and $\tau_{\varphi}$, as well as the spin-splitting energy $\Delta_{SO}$ are consistent with values measured in buried Ge QWs \cite{morrison2014observation,failla2015narrow,moriya2014cubic}. Finally, we find that weak anti-localization can as well be suppressed by an in-plane magnetic field, reflecting the finite thickness of the 2DHG and a contribution from Zeeman effect, surface roughness, and virtual inter-subband scattering processes.

The authors would like to thank M. Houzet, J. Meyer, Y.-M. Niquet, S. Oda, M. Sanquer and Z. Zeng for the fruitful discussions. We acknowledge financial support from the Agence Nationale de la Recherche, through the TOPONANO project, from the EU through the ERC grant No. 280043, and from the Fondation Nanosciences, Grenoble.


\begin{thebibliography}{41}%
\makeatletter
\providecommand \@ifxundefined [1]{%
 \@ifx{#1\undefined}
}%
\providecommand \@ifnum [1]{%
 \ifnum #1\expandafter \@firstoftwo
 \else \expandafter \@secondoftwo
 \fi
}%
\providecommand \@ifx [1]{%
 \ifx #1\expandafter \@firstoftwo
 \else \expandafter \@secondoftwo
 \fi
}%
\providecommand \natexlab [1]{#1}%
\providecommand \enquote  [1]{``#1''}%
\providecommand \bibnamefont  [1]{#1}%
\providecommand \bibfnamefont [1]{#1}%
\providecommand \citenamefont [1]{#1}%
\providecommand \href@noop [0]{\@secondoftwo}%
\providecommand \href [0]{\begingroup \@sanitize@url \@href}%
\providecommand \@href[1]{\@@startlink{#1}\@@href}%
\providecommand \@@href[1]{\endgroup#1\@@endlink}%
\providecommand \@sanitize@url [0]{\catcode `\\12\catcode `\$12\catcode
  `\&12\catcode `\#12\catcode `\^12\catcode `\_12\catcode `\%12\relax}%
\providecommand \@@startlink[1]{}%
\providecommand \@@endlink[0]{}%
\providecommand \url  [0]{\begingroup\@sanitize@url \@url }%
\providecommand \@url [1]{\endgroup\@href {#1}{\urlprefix }}%
\providecommand \urlprefix  [0]{URL }%
\providecommand \Eprint [0]{\href }%
\providecommand \doibase [0]{http://dx.doi.org/}%
\providecommand \selectlanguage [0]{\@gobble}%
\providecommand \bibinfo  [0]{\@secondoftwo}%
\providecommand \bibfield  [0]{\@secondoftwo}%
\providecommand \translation [1]{[#1]}%
\providecommand \BibitemOpen [0]{}%
\providecommand \bibitemStop [0]{}%
\providecommand \bibitemNoStop [0]{.\EOS\space}%
\providecommand \EOS [0]{\spacefactor3000\relax}%
\providecommand \BibitemShut  [1]{\csname bibitem#1\endcsname}%
\let\auto@bib@innerbib\@empty
\bibitem [{\citenamefont {Zwanenburg}\ \emph {et~al.}(2013)\citenamefont
  {Zwanenburg}, \citenamefont {Dzurak}, \citenamefont {Morello}, \citenamefont
  {Simmons}, \citenamefont {Hollenberg}, \citenamefont {Klimeck}, \citenamefont
  {Rogge}, \citenamefont {Coppersmith},\ and\ \citenamefont
  {Eriksson}}]{RevModPhys.85.961}%
  \BibitemOpen
  \bibfield  {author} {\bibinfo {author} {\bibfnamefont {F.~A.}\ \bibnamefont
  {Zwanenburg}}, \bibinfo {author} {\bibfnamefont {A.~S.}\ \bibnamefont
  {Dzurak}}, \bibinfo {author} {\bibfnamefont {A.}~\bibnamefont {Morello}},
  \bibinfo {author} {\bibfnamefont {M.~Y.}\ \bibnamefont {Simmons}}, \bibinfo
  {author} {\bibfnamefont {L.~C.~L.}\ \bibnamefont {Hollenberg}}, \bibinfo
  {author} {\bibfnamefont {G.}~\bibnamefont {Klimeck}}, \bibinfo {author}
  {\bibfnamefont {S.}~\bibnamefont {Rogge}}, \bibinfo {author} {\bibfnamefont
  {S.~N.}\ \bibnamefont {Coppersmith}}, \ and\ \bibinfo {author} {\bibfnamefont
  {M.~A.}\ \bibnamefont {Eriksson}},\ }\href {\doibase
  10.1103/RevModPhys.85.961} {\bibfield  {journal} {\bibinfo  {journal} {Rev.
  Mod. Phys.}\ }\textbf {\bibinfo {volume} {85}},\ \bibinfo {pages} {961}
  (\bibinfo {year} {2013})}\BibitemShut {NoStop}%
\bibitem [{\citenamefont {Morrison}\ and\ \citenamefont
  {Myronov}(2016)}]{PSSA:PSSA201600713}%
  \BibitemOpen
  \bibfield  {author} {\bibinfo {author} {\bibfnamefont {C.}~\bibnamefont
  {Morrison}}\ and\ \bibinfo {author} {\bibfnamefont {M.}~\bibnamefont
  {Myronov}},\ }\href {\doibase 10.1002/pssa.201600713} {\bibfield  {journal}
  {\bibinfo  {journal} {physica status solidi (a)}\ }\textbf {\bibinfo {volume}
  {213}},\ \bibinfo {pages} {2809} (\bibinfo {year} {2016})}\BibitemShut
  {NoStop}%
\bibitem [{\citenamefont {Sawano}\ \emph {et~al.}(2009)\citenamefont {Sawano},
  \citenamefont {Toyama}, \citenamefont {Masutomi}, \citenamefont {Okamoto},
  \citenamefont {Usami}, \citenamefont {Arimoto}, \citenamefont {Nakagawa},\
  and\ \citenamefont {Shiraki}}]{sawano2009strain}%
  \BibitemOpen
  \bibfield  {author} {\bibinfo {author} {\bibfnamefont {K.}~\bibnamefont
  {Sawano}}, \bibinfo {author} {\bibfnamefont {K.}~\bibnamefont {Toyama}},
  \bibinfo {author} {\bibfnamefont {R.}~\bibnamefont {Masutomi}}, \bibinfo
  {author} {\bibfnamefont {T.}~\bibnamefont {Okamoto}}, \bibinfo {author}
  {\bibfnamefont {N.}~\bibnamefont {Usami}}, \bibinfo {author} {\bibfnamefont
  {K.}~\bibnamefont {Arimoto}}, \bibinfo {author} {\bibfnamefont
  {K.}~\bibnamefont {Nakagawa}}, \ and\ \bibinfo {author} {\bibfnamefont
  {Y.}~\bibnamefont {Shiraki}},\ }\href@noop {} {\bibfield  {journal} {\bibinfo
   {journal} {Applied Physics Letters}\ }\textbf {\bibinfo {volume} {95}},\
  \bibinfo {pages} {2109} (\bibinfo {year} {2009})}\BibitemShut {NoStop}%
\bibitem [{\citenamefont {Morrison}\ \emph
  {et~al.}(2016{\natexlab{a}})\citenamefont {Morrison}, \citenamefont
  {Casteleiro}, \citenamefont {Leadley},\ and\ \citenamefont
  {Myronov}}]{doi:10.1063/1.4962432}%
  \BibitemOpen
  \bibfield  {author} {\bibinfo {author} {\bibfnamefont {C.}~\bibnamefont
  {Morrison}}, \bibinfo {author} {\bibfnamefont {C.}~\bibnamefont
  {Casteleiro}}, \bibinfo {author} {\bibfnamefont {D.~R.}\ \bibnamefont
  {Leadley}}, \ and\ \bibinfo {author} {\bibfnamefont {M.}~\bibnamefont
  {Myronov}},\ }\href {\doibase 10.1063/1.4962432} {\bibfield  {journal}
  {\bibinfo  {journal} {Applied Physics Letters}\ }\textbf {\bibinfo {volume}
  {109}},\ \bibinfo {pages} {102103} (\bibinfo {year} {2016}{\natexlab{a}})},\
  \Eprint {http://arxiv.org/abs/http://dx.doi.org/10.1063/1.4962432}
  {http://dx.doi.org/10.1063/1.4962432} \BibitemShut {NoStop}%
\bibitem [{\citenamefont {Higginbotham}\ \emph {et~al.}(2014)\citenamefont
  {Higginbotham}, \citenamefont {Larsen}, \citenamefont {Yao}, \citenamefont
  {Yan}, \citenamefont {Lieber}, \citenamefont {Marcus},\ and\ \citenamefont
  {Kuemmeth}}]{doi:10.1021/nl501242b}%
  \BibitemOpen
  \bibfield  {author} {\bibinfo {author} {\bibfnamefont {A.~P.}\ \bibnamefont
  {Higginbotham}}, \bibinfo {author} {\bibfnamefont {T.~W.}\ \bibnamefont
  {Larsen}}, \bibinfo {author} {\bibfnamefont {J.}~\bibnamefont {Yao}},
  \bibinfo {author} {\bibfnamefont {H.}~\bibnamefont {Yan}}, \bibinfo {author}
  {\bibfnamefont {C.~M.}\ \bibnamefont {Lieber}}, \bibinfo {author}
  {\bibfnamefont {C.~M.}\ \bibnamefont {Marcus}}, \ and\ \bibinfo {author}
  {\bibfnamefont {F.}~\bibnamefont {Kuemmeth}},\ }\href {\doibase
  10.1021/nl501242b} {\bibfield  {journal} {\bibinfo  {journal} {Nano Letters}\
  }\textbf {\bibinfo {volume} {14}},\ \bibinfo {pages} {3582} (\bibinfo {year}
  {2014})}\BibitemShut {NoStop}%
\bibitem [{\citenamefont {Testelin}\ \emph {et~al.}(2009)\citenamefont
  {Testelin}, \citenamefont {Bernardot}, \citenamefont {Eble},\ and\
  \citenamefont {Chamarro}}]{testelin2009hole}%
  \BibitemOpen
  \bibfield  {author} {\bibinfo {author} {\bibfnamefont {C.}~\bibnamefont
  {Testelin}}, \bibinfo {author} {\bibfnamefont {F.}~\bibnamefont {Bernardot}},
  \bibinfo {author} {\bibfnamefont {B.}~\bibnamefont {Eble}}, \ and\ \bibinfo
  {author} {\bibfnamefont {M.}~\bibnamefont {Chamarro}},\ }\href@noop {}
  {\bibfield  {journal} {\bibinfo  {journal} {Physical Review B}\ }\textbf
  {\bibinfo {volume} {79}},\ \bibinfo {pages} {195440} (\bibinfo {year}
  {2009})}\BibitemShut {NoStop}%
\bibitem [{\citenamefont {Watzinger}\ \emph {et~al.}(2016)\citenamefont
  {Watzinger}, \citenamefont {Kloeffel}, \citenamefont {Vuku\v{s}i\'{c}},
  \citenamefont {Rossell}, \citenamefont {Sessi}, \citenamefont {Kuku\v{c}ka},
  \citenamefont {Kirchschlager}, \citenamefont {Lausecker}, \citenamefont
  {Truhlar}, \citenamefont {Glaser}, \citenamefont {Rastelli}, \citenamefont
  {Fuhrer}, \citenamefont {Loss},\ and\ \citenamefont
  {Katsaros}}]{doi:10.1021/acs.nanolett.6b02715}%
  \BibitemOpen
  \bibfield  {author} {\bibinfo {author} {\bibfnamefont {H.}~\bibnamefont
  {Watzinger}}, \bibinfo {author} {\bibfnamefont {C.}~\bibnamefont {Kloeffel}},
  \bibinfo {author} {\bibfnamefont {L.}~\bibnamefont {Vuku\v{s}i\'{c}}},
  \bibinfo {author} {\bibfnamefont {M.~D.}\ \bibnamefont {Rossell}}, \bibinfo
  {author} {\bibfnamefont {V.}~\bibnamefont {Sessi}}, \bibinfo {author}
  {\bibfnamefont {J.}~\bibnamefont {Kuku\v{c}ka}}, \bibinfo {author}
  {\bibfnamefont {R.}~\bibnamefont {Kirchschlager}}, \bibinfo {author}
  {\bibfnamefont {E.}~\bibnamefont {Lausecker}}, \bibinfo {author}
  {\bibfnamefont {A.}~\bibnamefont {Truhlar}}, \bibinfo {author} {\bibfnamefont
  {M.}~\bibnamefont {Glaser}}, \bibinfo {author} {\bibfnamefont
  {A.}~\bibnamefont {Rastelli}}, \bibinfo {author} {\bibfnamefont
  {A.}~\bibnamefont {Fuhrer}}, \bibinfo {author} {\bibfnamefont
  {D.}~\bibnamefont {Loss}}, \ and\ \bibinfo {author} {\bibfnamefont
  {G.}~\bibnamefont {Katsaros}},\ }\href {\doibase
  10.1021/acs.nanolett.6b02715} {\bibfield  {journal} {\bibinfo  {journal}
  {Nano Letters}\ }\textbf {\bibinfo {volume} {16}},\ \bibinfo {pages} {6879}
  (\bibinfo {year} {2016})},\ \bibinfo {note} {pMID: 27656760},\ \Eprint
  {http://arxiv.org/abs/http://dx.doi.org/10.1021/acs.nanolett.6b02715}
  {http://dx.doi.org/10.1021/acs.nanolett.6b02715} \BibitemShut {NoStop}%
\bibitem [{\citenamefont {Ares}\ \emph
  {et~al.}(2013{\natexlab{a}})\citenamefont {Ares}, \citenamefont {Golovach},
  \citenamefont {Katsaros}, \citenamefont {Stoffel}, \citenamefont {Fournel},
  \citenamefont {Glazman}, \citenamefont {Schmidt},\ and\ \citenamefont
  {De~Franceschi}}]{PhysRevLett.110.046602}%
  \BibitemOpen
  \bibfield  {author} {\bibinfo {author} {\bibfnamefont {N.}~\bibnamefont
  {Ares}}, \bibinfo {author} {\bibfnamefont {V.~N.}\ \bibnamefont {Golovach}},
  \bibinfo {author} {\bibfnamefont {G.}~\bibnamefont {Katsaros}}, \bibinfo
  {author} {\bibfnamefont {M.}~\bibnamefont {Stoffel}}, \bibinfo {author}
  {\bibfnamefont {F.}~\bibnamefont {Fournel}}, \bibinfo {author} {\bibfnamefont
  {L.~I.}\ \bibnamefont {Glazman}}, \bibinfo {author} {\bibfnamefont {O.~G.}\
  \bibnamefont {Schmidt}}, \ and\ \bibinfo {author} {\bibfnamefont
  {S.}~\bibnamefont {De~Franceschi}},\ }\href {\doibase
  10.1103/PhysRevLett.110.046602} {\bibfield  {journal} {\bibinfo  {journal}
  {Phys. Rev. Lett.}\ }\textbf {\bibinfo {volume} {110}},\ \bibinfo {pages}
  {046602} (\bibinfo {year} {2013}{\natexlab{a}})}\BibitemShut {NoStop}%
\bibitem [{\citenamefont {Kloeffel}\ \emph {et~al.}(2011)\citenamefont
  {Kloeffel}, \citenamefont {Trif},\ and\ \citenamefont
  {Loss}}]{PhysRevB.84.195314}%
  \BibitemOpen
  \bibfield  {author} {\bibinfo {author} {\bibfnamefont {C.}~\bibnamefont
  {Kloeffel}}, \bibinfo {author} {\bibfnamefont {M.}~\bibnamefont {Trif}}, \
  and\ \bibinfo {author} {\bibfnamefont {D.}~\bibnamefont {Loss}},\ }\href
  {\doibase 10.1103/PhysRevB.84.195314} {\bibfield  {journal} {\bibinfo
  {journal} {Phys. Rev. B}\ }\textbf {\bibinfo {volume} {84}},\ \bibinfo
  {pages} {195314} (\bibinfo {year} {2011})}\BibitemShut {NoStop}%
\bibitem [{\citenamefont {Morrison}\ \emph
  {et~al.}(2014{\natexlab{a}})\citenamefont {Morrison}, \citenamefont
  {Wi{\'s}niewski}, \citenamefont {Rhead}, \citenamefont {Foronda},
  \citenamefont {Leadley},\ and\ \citenamefont
  {Myronov}}]{doi:10.1063/1.4901107}%
  \BibitemOpen
  \bibfield  {author} {\bibinfo {author} {\bibfnamefont {C.}~\bibnamefont
  {Morrison}}, \bibinfo {author} {\bibfnamefont {P.}~\bibnamefont
  {Wi{\'s}niewski}}, \bibinfo {author} {\bibfnamefont {S.~D.}\ \bibnamefont
  {Rhead}}, \bibinfo {author} {\bibfnamefont {J.}~\bibnamefont {Foronda}},
  \bibinfo {author} {\bibfnamefont {D.~R.}\ \bibnamefont {Leadley}}, \ and\
  \bibinfo {author} {\bibfnamefont {M.}~\bibnamefont {Myronov}},\ }\href
  {\doibase 10.1063/1.4901107} {\bibfield  {journal} {\bibinfo  {journal}
  {Applied Physics Letters}\ }\textbf {\bibinfo {volume} {105}},\ \bibinfo
  {pages} {182401} (\bibinfo {year} {2014}{\natexlab{a}})},\ \Eprint
  {http://arxiv.org/abs/http://dx.doi.org/10.1063/1.4901107}
  {http://dx.doi.org/10.1063/1.4901107} \BibitemShut {NoStop}%
\bibitem [{\citenamefont {Zarassi}\ \emph {et~al.}(2016)\citenamefont
  {Zarassi}, \citenamefont {Su}, \citenamefont {Danon}, \citenamefont
  {Schwenderling}, \citenamefont {Hocevar}, \citenamefont {Nguyen},
  \citenamefont {Yoo}, \citenamefont {Dayeh},\ and\ \citenamefont
  {Frolov}}]{zarassi2016magnetic}%
  \BibitemOpen
  \bibfield  {author} {\bibinfo {author} {\bibfnamefont {A.}~\bibnamefont
  {Zarassi}}, \bibinfo {author} {\bibfnamefont {Z.}~\bibnamefont {Su}},
  \bibinfo {author} {\bibfnamefont {J.}~\bibnamefont {Danon}}, \bibinfo
  {author} {\bibfnamefont {J.}~\bibnamefont {Schwenderling}}, \bibinfo {author}
  {\bibfnamefont {M.}~\bibnamefont {Hocevar}}, \bibinfo {author} {\bibfnamefont
  {B.-M.}\ \bibnamefont {Nguyen}}, \bibinfo {author} {\bibfnamefont
  {J.}~\bibnamefont {Yoo}}, \bibinfo {author} {\bibfnamefont {S.~A.}\
  \bibnamefont {Dayeh}}, \ and\ \bibinfo {author} {\bibfnamefont {S.~M.}\
  \bibnamefont {Frolov}},\ }\href@noop {} {\bibfield  {journal} {\bibinfo
  {journal} {arXiv preprint arXiv:1610.04596}\ } (\bibinfo {year}
  {2016})}\BibitemShut {NoStop}%
\bibitem [{\citenamefont {Ares}\ \emph
  {et~al.}(2013{\natexlab{b}})\citenamefont {Ares}, \citenamefont {Katsaros},
  \citenamefont {Golovach}, \citenamefont {Zhang}, \citenamefont {Prager},
  \citenamefont {Glazman}, \citenamefont {Schmidt},\ and\ \citenamefont
  {Franceschi}}]{doi:10.1063/1.4858959}%
  \BibitemOpen
  \bibfield  {author} {\bibinfo {author} {\bibfnamefont {N.}~\bibnamefont
  {Ares}}, \bibinfo {author} {\bibfnamefont {G.}~\bibnamefont {Katsaros}},
  \bibinfo {author} {\bibfnamefont {V.~N.}\ \bibnamefont {Golovach}}, \bibinfo
  {author} {\bibfnamefont {J.~J.}\ \bibnamefont {Zhang}}, \bibinfo {author}
  {\bibfnamefont {A.}~\bibnamefont {Prager}}, \bibinfo {author} {\bibfnamefont
  {L.~I.}\ \bibnamefont {Glazman}}, \bibinfo {author} {\bibfnamefont {O.~G.}\
  \bibnamefont {Schmidt}}, \ and\ \bibinfo {author} {\bibfnamefont {S.~D.}\
  \bibnamefont {Franceschi}},\ }\href {\doibase 10.1063/1.4858959} {\bibfield
  {journal} {\bibinfo  {journal} {Applied Physics Letters}\ }\textbf {\bibinfo
  {volume} {103}},\ \bibinfo {pages} {263113} (\bibinfo {year}
  {2013}{\natexlab{b}})}\BibitemShut {NoStop}%
\bibitem [{\citenamefont {Brauns}\ \emph {et~al.}(2016)\citenamefont {Brauns},
  \citenamefont {Ridderbos}, \citenamefont {Li}, \citenamefont {van~der Wiel},
  \citenamefont {Bakkers},\ and\ \citenamefont
  {Zwanenburg}}]{doi:10.1063/1.4963715}%
  \BibitemOpen
  \bibfield  {author} {\bibinfo {author} {\bibfnamefont {M.}~\bibnamefont
  {Brauns}}, \bibinfo {author} {\bibfnamefont {J.}~\bibnamefont {Ridderbos}},
  \bibinfo {author} {\bibfnamefont {A.}~\bibnamefont {Li}}, \bibinfo {author}
  {\bibfnamefont {W.~G.}\ \bibnamefont {van~der Wiel}}, \bibinfo {author}
  {\bibfnamefont {E.~P. A.~M.}\ \bibnamefont {Bakkers}}, \ and\ \bibinfo
  {author} {\bibfnamefont {F.~A.}\ \bibnamefont {Zwanenburg}},\ }\href
  {\doibase 10.1063/1.4963715} {\bibfield  {journal} {\bibinfo  {journal}
  {Applied Physics Letters}\ }\textbf {\bibinfo {volume} {109}},\ \bibinfo
  {pages} {143113} (\bibinfo {year} {2016})}\BibitemShut {NoStop}%
\bibitem [{\citenamefont {Maurand}\ \emph {et~al.}(2016)\citenamefont
  {Maurand}, \citenamefont {Jehl}, \citenamefont {Kotekar-Patil}, \citenamefont
  {Corna}, \citenamefont {Bohuslavskyi}, \citenamefont {Lavi{\'e}ville},
  \citenamefont {Hutin}, \citenamefont {Barraud}, \citenamefont {Vinet},
  \citenamefont {Sanquer} \emph {et~al.}}]{maurand2016cmos}%
  \BibitemOpen
  \bibfield  {author} {\bibinfo {author} {\bibfnamefont {R.}~\bibnamefont
  {Maurand}}, \bibinfo {author} {\bibfnamefont {X.}~\bibnamefont {Jehl}},
  \bibinfo {author} {\bibfnamefont {D.}~\bibnamefont {Kotekar-Patil}}, \bibinfo
  {author} {\bibfnamefont {A.}~\bibnamefont {Corna}}, \bibinfo {author}
  {\bibfnamefont {H.}~\bibnamefont {Bohuslavskyi}}, \bibinfo {author}
  {\bibfnamefont {R.}~\bibnamefont {Lavi{\'e}ville}}, \bibinfo {author}
  {\bibfnamefont {L.}~\bibnamefont {Hutin}}, \bibinfo {author} {\bibfnamefont
  {S.}~\bibnamefont {Barraud}}, \bibinfo {author} {\bibfnamefont
  {M.}~\bibnamefont {Vinet}}, \bibinfo {author} {\bibfnamefont
  {M.}~\bibnamefont {Sanquer}},  \emph {et~al.},\ }\href@noop {} {\bibfield
  {journal} {\bibinfo  {journal} {Nature Communications}\ }\textbf {\bibinfo
  {volume} {7}},\ \bibinfo {pages} {13575} (\bibinfo {year}
  {2016})}\BibitemShut {NoStop}%
\bibitem [{\citenamefont {Maier}\ \emph {et~al.}(2014)\citenamefont {Maier},
  \citenamefont {Klinovaja},\ and\ \citenamefont {Loss}}]{PhysRevB.90.195421}%
  \BibitemOpen
  \bibfield  {author} {\bibinfo {author} {\bibfnamefont {F.}~\bibnamefont
  {Maier}}, \bibinfo {author} {\bibfnamefont {J.}~\bibnamefont {Klinovaja}}, \
  and\ \bibinfo {author} {\bibfnamefont {D.}~\bibnamefont {Loss}},\ }\href
  {\doibase 10.1103/PhysRevB.90.195421} {\bibfield  {journal} {\bibinfo
  {journal} {Phys. Rev. B}\ }\textbf {\bibinfo {volume} {90}},\ \bibinfo
  {pages} {195421} (\bibinfo {year} {2014})}\BibitemShut {NoStop}%
\bibitem [{\citenamefont {Su}\ \emph {et~al.}(2016)\citenamefont {Su},
  \citenamefont {Zarassi}, \citenamefont {Nguyen}, \citenamefont {Yoo},
  \citenamefont {Dayeh},\ and\ \citenamefont {Frolov}}]{su2016high}%
  \BibitemOpen
  \bibfield  {author} {\bibinfo {author} {\bibfnamefont {Z.}~\bibnamefont
  {Su}}, \bibinfo {author} {\bibfnamefont {A.}~\bibnamefont {Zarassi}},
  \bibinfo {author} {\bibfnamefont {B.-M.}\ \bibnamefont {Nguyen}}, \bibinfo
  {author} {\bibfnamefont {J.}~\bibnamefont {Yoo}}, \bibinfo {author}
  {\bibfnamefont {S.~A.}\ \bibnamefont {Dayeh}}, \ and\ \bibinfo {author}
  {\bibfnamefont {S.~M.}\ \bibnamefont {Frolov}},\ }\href@noop {} {\bibfield
  {journal} {\bibinfo  {journal} {arXiv preprint arXiv:1610.03010}\ } (\bibinfo
  {year} {2016})}\BibitemShut {NoStop}%
\bibitem [{\citenamefont {Myronov}\ \emph {et~al.}(2010)\citenamefont
  {Myronov}, \citenamefont {Dobbie}, \citenamefont {Shah}, \citenamefont {Liu},
  \citenamefont {Nguyen},\ and\ \citenamefont {Leadley}}]{myronov2010high}%
  \BibitemOpen
  \bibfield  {author} {\bibinfo {author} {\bibfnamefont {M.}~\bibnamefont
  {Myronov}}, \bibinfo {author} {\bibfnamefont {A.}~\bibnamefont {Dobbie}},
  \bibinfo {author} {\bibfnamefont {V.~A.}\ \bibnamefont {Shah}}, \bibinfo
  {author} {\bibfnamefont {X.-C.}\ \bibnamefont {Liu}}, \bibinfo {author}
  {\bibfnamefont {V.~H.}\ \bibnamefont {Nguyen}}, \ and\ \bibinfo {author}
  {\bibfnamefont {D.~R.}\ \bibnamefont {Leadley}},\ }\href@noop {} {\bibfield
  {journal} {\bibinfo  {journal} {Electrochemical and Solid-State Letters}\
  }\textbf {\bibinfo {volume} {13}},\ \bibinfo {pages} {H388} (\bibinfo {year}
  {2010})}\BibitemShut {NoStop}%
\bibitem [{\citenamefont {Dobbie}\ \emph
  {et~al.}(2012{\natexlab{a}})\citenamefont {Dobbie}, \citenamefont {Myronov},
  \citenamefont {Morris}, \citenamefont {Hassan}, \citenamefont {Prest},
  \citenamefont {Shah}, \citenamefont {Parker}, \citenamefont {Whall},\ and\
  \citenamefont {Leadley}}]{doi:10.1063/1.4763476}%
  \BibitemOpen
  \bibfield  {author} {\bibinfo {author} {\bibfnamefont {A.}~\bibnamefont
  {Dobbie}}, \bibinfo {author} {\bibfnamefont {M.}~\bibnamefont {Myronov}},
  \bibinfo {author} {\bibfnamefont {R.~J.~H.}\ \bibnamefont {Morris}}, \bibinfo
  {author} {\bibfnamefont {A.~H.~A.}\ \bibnamefont {Hassan}}, \bibinfo {author}
  {\bibfnamefont {M.~J.}\ \bibnamefont {Prest}}, \bibinfo {author}
  {\bibfnamefont {V.~A.}\ \bibnamefont {Shah}}, \bibinfo {author}
  {\bibfnamefont {E.~H.~C.}\ \bibnamefont {Parker}}, \bibinfo {author}
  {\bibfnamefont {T.~E.}\ \bibnamefont {Whall}}, \ and\ \bibinfo {author}
  {\bibfnamefont {D.~R.}\ \bibnamefont {Leadley}},\ }\href {\doibase
  10.1063/1.4763476} {\bibfield  {journal} {\bibinfo  {journal} {Applied
  Physics Letters}\ }\textbf {\bibinfo {volume} {101}},\ \bibinfo {pages}
  {172108} (\bibinfo {year} {2012}{\natexlab{a}})},\ \Eprint
  {http://arxiv.org/abs/http://dx.doi.org/10.1063/1.4763476}
  {http://dx.doi.org/10.1063/1.4763476} \BibitemShut {NoStop}%
\bibitem [{\citenamefont {Myronov}\ \emph
  {et~al.}(2014{\natexlab{a}})\citenamefont {Myronov}, \citenamefont
  {Morrison}, \citenamefont {Halpin}, \citenamefont {Rhead}, \citenamefont
  {Casteleiro}, \citenamefont {Foronda}, \citenamefont {Shah},\ and\
  \citenamefont {Leadley}}]{1347-4065-53-4S-04EH02}%
  \BibitemOpen
  \bibfield  {author} {\bibinfo {author} {\bibfnamefont {M.}~\bibnamefont
  {Myronov}}, \bibinfo {author} {\bibfnamefont {C.}~\bibnamefont {Morrison}},
  \bibinfo {author} {\bibfnamefont {J.}~\bibnamefont {Halpin}}, \bibinfo
  {author} {\bibfnamefont {S.}~\bibnamefont {Rhead}}, \bibinfo {author}
  {\bibfnamefont {C.}~\bibnamefont {Casteleiro}}, \bibinfo {author}
  {\bibfnamefont {J.}~\bibnamefont {Foronda}}, \bibinfo {author} {\bibfnamefont
  {V.~A.}\ \bibnamefont {Shah}}, \ and\ \bibinfo {author} {\bibfnamefont
  {D.}~\bibnamefont {Leadley}},\ }\href
  {http://stacks.iop.org/1347-4065/53/i=4S/a=04EH02} {\bibfield  {journal}
  {\bibinfo  {journal} {Japanese Journal of Applied Physics}\ }\textbf
  {\bibinfo {volume} {53}},\ \bibinfo {pages} {04EH02} (\bibinfo {year}
  {2014}{\natexlab{a}})}\BibitemShut {NoStop}%
\bibitem [{\citenamefont {Myronov}\ \emph
  {et~al.}(2014{\natexlab{b}})\citenamefont {Myronov}, \citenamefont
  {Morrison}, \citenamefont {Halpin}, \citenamefont {Rhead}, \citenamefont
  {Foronda},\ and\ \citenamefont {Leadley}}]{6874628}%
  \BibitemOpen
  \bibfield  {author} {\bibinfo {author} {\bibfnamefont {M.}~\bibnamefont
  {Myronov}}, \bibinfo {author} {\bibfnamefont {C.}~\bibnamefont {Morrison}},
  \bibinfo {author} {\bibfnamefont {J.}~\bibnamefont {Halpin}}, \bibinfo
  {author} {\bibfnamefont {S.}~\bibnamefont {Rhead}}, \bibinfo {author}
  {\bibfnamefont {J.}~\bibnamefont {Foronda}}, \ and\ \bibinfo {author}
  {\bibfnamefont {D.}~\bibnamefont {Leadley}},\ }in\ \href {\doibase
  10.1109/ISTDM.2014.6874628} {\emph {\bibinfo {booktitle} {2014 7th
  International Silicon-Germanium Technology and Device Meeting (ISTDM)}}}\
  (\bibinfo {year} {2014})\ pp.\ \bibinfo {pages} {11--12}\BibitemShut
  {NoStop}%
\bibitem [{\citenamefont {Shi}\ \emph {et~al.}(2015)\citenamefont {Shi},
  \citenamefont {Zudov}, \citenamefont {Morrison},\ and\ \citenamefont
  {Myronov}}]{PhysRevB.91.241303}%
  \BibitemOpen
  \bibfield  {author} {\bibinfo {author} {\bibfnamefont {Q.}~\bibnamefont
  {Shi}}, \bibinfo {author} {\bibfnamefont {M.~A.}\ \bibnamefont {Zudov}},
  \bibinfo {author} {\bibfnamefont {C.}~\bibnamefont {Morrison}}, \ and\
  \bibinfo {author} {\bibfnamefont {M.}~\bibnamefont {Myronov}},\ }\href
  {\doibase 10.1103/PhysRevB.91.241303} {\bibfield  {journal} {\bibinfo
  {journal} {Phys. Rev. B}\ }\textbf {\bibinfo {volume} {91}},\ \bibinfo
  {pages} {241303} (\bibinfo {year} {2015})}\BibitemShut {NoStop}%
\bibitem [{\citenamefont {Morrison}\ \emph
  {et~al.}(2014{\natexlab{b}})\citenamefont {Morrison}, \citenamefont
  {Wi{\'s}niewski}, \citenamefont {Rhead}, \citenamefont {Foronda},
  \citenamefont {Leadley},\ and\ \citenamefont
  {Myronov}}]{morrison2014observation}%
  \BibitemOpen
  \bibfield  {author} {\bibinfo {author} {\bibfnamefont {C.}~\bibnamefont
  {Morrison}}, \bibinfo {author} {\bibfnamefont {P.}~\bibnamefont
  {Wi{\'s}niewski}}, \bibinfo {author} {\bibfnamefont {S.}~\bibnamefont
  {Rhead}}, \bibinfo {author} {\bibfnamefont {J.}~\bibnamefont {Foronda}},
  \bibinfo {author} {\bibfnamefont {D.~R.}\ \bibnamefont {Leadley}}, \ and\
  \bibinfo {author} {\bibfnamefont {M.}~\bibnamefont {Myronov}},\ }\href@noop
  {} {\bibfield  {journal} {\bibinfo  {journal} {Applied Physics Letters}\
  }\textbf {\bibinfo {volume} {105}},\ \bibinfo {pages} {182401} (\bibinfo
  {year} {2014}{\natexlab{b}})}\BibitemShut {NoStop}%
\bibitem [{\citenamefont {Failla}\ \emph {et~al.}(2015)\citenamefont {Failla},
  \citenamefont {Myronov}, \citenamefont {Morrison}, \citenamefont {Leadley},\
  and\ \citenamefont {Lloyd-Hughes}}]{failla2015narrow}%
  \BibitemOpen
  \bibfield  {author} {\bibinfo {author} {\bibfnamefont {M.}~\bibnamefont
  {Failla}}, \bibinfo {author} {\bibfnamefont {M.}~\bibnamefont {Myronov}},
  \bibinfo {author} {\bibfnamefont {C.}~\bibnamefont {Morrison}}, \bibinfo
  {author} {\bibfnamefont {D.}~\bibnamefont {Leadley}}, \ and\ \bibinfo
  {author} {\bibfnamefont {J.}~\bibnamefont {Lloyd-Hughes}},\ }\href@noop {}
  {\bibfield  {journal} {\bibinfo  {journal} {Physical Review B}\ }\textbf
  {\bibinfo {volume} {92}},\ \bibinfo {pages} {045303} (\bibinfo {year}
  {2015})}\BibitemShut {NoStop}%
\bibitem [{\citenamefont {Morrison}\ \emph
  {et~al.}(2016{\natexlab{b}})\citenamefont {Morrison}, \citenamefont
  {Foronda}, \citenamefont {Wi{\'s}niewski}, \citenamefont {Rhead},
  \citenamefont {Leadley},\ and\ \citenamefont {Myronov}}]{Morrison201684}%
  \BibitemOpen
  \bibfield  {author} {\bibinfo {author} {\bibfnamefont {C.}~\bibnamefont
  {Morrison}}, \bibinfo {author} {\bibfnamefont {J.}~\bibnamefont {Foronda}},
  \bibinfo {author} {\bibfnamefont {P.}~\bibnamefont {Wi{\'s}niewski}},
  \bibinfo {author} {\bibfnamefont {S.}~\bibnamefont {Rhead}}, \bibinfo
  {author} {\bibfnamefont {D.}~\bibnamefont {Leadley}}, \ and\ \bibinfo
  {author} {\bibfnamefont {M.}~\bibnamefont {Myronov}},\ }\href {\doibase
  http://dx.doi.org/10.1016/j.tsf.2015.09.063} {\bibfield  {journal} {\bibinfo
  {journal} {Thin Solid Films}\ }\textbf {\bibinfo {volume} {602}},\ \bibinfo
  {pages} {84 } (\bibinfo {year} {2016}{\natexlab{b}})},\ \bibinfo {note} {the
  9th International Conference on Silicon Epitaxy and
  Heterostructures}\BibitemShut {NoStop}%
\bibitem [{\citenamefont {Foronda}\ \emph {et~al.}(2015)\citenamefont
  {Foronda}, \citenamefont {Morrison}, \citenamefont {Halpin}, \citenamefont
  {Rhead},\ and\ \citenamefont {Myronov}}]{0953-8984-27-2-022201}%
  \BibitemOpen
  \bibfield  {author} {\bibinfo {author} {\bibfnamefont {J.}~\bibnamefont
  {Foronda}}, \bibinfo {author} {\bibfnamefont {C.}~\bibnamefont {Morrison}},
  \bibinfo {author} {\bibfnamefont {J.~E.}\ \bibnamefont {Halpin}}, \bibinfo
  {author} {\bibfnamefont {S.~D.}\ \bibnamefont {Rhead}}, \ and\ \bibinfo
  {author} {\bibfnamefont {M.}~\bibnamefont {Myronov}},\ }\href
  {http://stacks.iop.org/0953-8984/27/i=2/a=022201} {\bibfield  {journal}
  {\bibinfo  {journal} {Journal of Physics: Condensed Matter}\ }\textbf
  {\bibinfo {volume} {27}},\ \bibinfo {pages} {022201} (\bibinfo {year}
  {2015})}\BibitemShut {NoStop}%
\bibitem [{\citenamefont {Failla}\ \emph {et~al.}(2016)\citenamefont {Failla},
  \citenamefont {Keller}, \citenamefont {Scalari}, \citenamefont {Maissen},
  \citenamefont {Faist}, \citenamefont {Reichl}, \citenamefont {Wegscheider},
  \citenamefont {Newell}, \citenamefont {Leadley}, \citenamefont {Myronov},\
  and\ \citenamefont {Lloyd-Hughes}}]{1367-2630-18-11-113036}%
  \BibitemOpen
  \bibfield  {author} {\bibinfo {author} {\bibfnamefont {M.}~\bibnamefont
  {Failla}}, \bibinfo {author} {\bibfnamefont {J.}~\bibnamefont {Keller}},
  \bibinfo {author} {\bibfnamefont {G.}~\bibnamefont {Scalari}}, \bibinfo
  {author} {\bibfnamefont {C.}~\bibnamefont {Maissen}}, \bibinfo {author}
  {\bibfnamefont {J.}~\bibnamefont {Faist}}, \bibinfo {author} {\bibfnamefont
  {C.}~\bibnamefont {Reichl}}, \bibinfo {author} {\bibfnamefont
  {W.}~\bibnamefont {Wegscheider}}, \bibinfo {author} {\bibfnamefont {O.~J.}\
  \bibnamefont {Newell}}, \bibinfo {author} {\bibfnamefont {D.~R.}\
  \bibnamefont {Leadley}}, \bibinfo {author} {\bibfnamefont {M.}~\bibnamefont
  {Myronov}}, \ and\ \bibinfo {author} {\bibfnamefont {J.}~\bibnamefont
  {Lloyd-Hughes}},\ }\href {http://stacks.iop.org/1367-2630/18/i=11/a=113036}
  {\bibfield  {journal} {\bibinfo  {journal} {New Journal of Physics}\ }\textbf
  {\bibinfo {volume} {18}},\ \bibinfo {pages} {113036} (\bibinfo {year}
  {2016})}\BibitemShut {NoStop}%
\bibitem [{\citenamefont {Dobbie}\ \emph
  {et~al.}(2012{\natexlab{b}})\citenamefont {Dobbie}, \citenamefont {Myronov},
  \citenamefont {Morris}, \citenamefont {Hassan}, \citenamefont {Prest},
  \citenamefont {Shah}, \citenamefont {Parker}, \citenamefont {Whall},\ and\
  \citenamefont {Leadley}}]{dobbie2012ultra}%
  \BibitemOpen
  \bibfield  {author} {\bibinfo {author} {\bibfnamefont {A.}~\bibnamefont
  {Dobbie}}, \bibinfo {author} {\bibfnamefont {M.}~\bibnamefont {Myronov}},
  \bibinfo {author} {\bibfnamefont {R.~J.}\ \bibnamefont {Morris}}, \bibinfo
  {author} {\bibfnamefont {A.}~\bibnamefont {Hassan}}, \bibinfo {author}
  {\bibfnamefont {M.~J.}\ \bibnamefont {Prest}}, \bibinfo {author}
  {\bibfnamefont {V.}~\bibnamefont {Shah}}, \bibinfo {author} {\bibfnamefont
  {E.~H.}\ \bibnamefont {Parker}}, \bibinfo {author} {\bibfnamefont {T.~E.}\
  \bibnamefont {Whall}}, \ and\ \bibinfo {author} {\bibfnamefont {D.~R.}\
  \bibnamefont {Leadley}},\ }\href@noop {} {\bibfield  {journal} {\bibinfo
  {journal} {Applied Physics Letters}\ }\textbf {\bibinfo {volume} {101}},\
  \bibinfo {pages} {172108} (\bibinfo {year} {2012}{\natexlab{b}})}\BibitemShut
  {NoStop}%
\bibitem [{\citenamefont {Knap}\ \emph {et~al.}(1996)\citenamefont {Knap},
  \citenamefont {Skierbiszewski}, \citenamefont {Zduniak}, \citenamefont
  {Litwin-Staszewska}, \citenamefont {Bertho}, \citenamefont {Kobbi},
  \citenamefont {Robert}, \citenamefont {Pikus}, \citenamefont {Pikus},
  \citenamefont {Iordanskii}, \citenamefont {Mosser}, \citenamefont
  {Zekentes},\ and\ \citenamefont {Lyanda-Geller}}]{PhysRevB.53.3912}%
  \BibitemOpen
  \bibfield  {author} {\bibinfo {author} {\bibfnamefont {W.}~\bibnamefont
  {Knap}}, \bibinfo {author} {\bibfnamefont {C.}~\bibnamefont
  {Skierbiszewski}}, \bibinfo {author} {\bibfnamefont {A.}~\bibnamefont
  {Zduniak}}, \bibinfo {author} {\bibfnamefont {E.}~\bibnamefont
  {Litwin-Staszewska}}, \bibinfo {author} {\bibfnamefont {D.}~\bibnamefont
  {Bertho}}, \bibinfo {author} {\bibfnamefont {F.}~\bibnamefont {Kobbi}},
  \bibinfo {author} {\bibfnamefont {J.~L.}\ \bibnamefont {Robert}}, \bibinfo
  {author} {\bibfnamefont {G.~E.}\ \bibnamefont {Pikus}}, \bibinfo {author}
  {\bibfnamefont {F.~G.}\ \bibnamefont {Pikus}}, \bibinfo {author}
  {\bibfnamefont {S.~V.}\ \bibnamefont {Iordanskii}}, \bibinfo {author}
  {\bibfnamefont {V.}~\bibnamefont {Mosser}}, \bibinfo {author} {\bibfnamefont
  {K.}~\bibnamefont {Zekentes}}, \ and\ \bibinfo {author} {\bibfnamefont
  {Y.~B.}\ \bibnamefont {Lyanda-Geller}},\ }\href {\doibase
  10.1103/PhysRevB.53.3912} {\bibfield  {journal} {\bibinfo  {journal} {Phys.
  Rev. B}\ }\textbf {\bibinfo {volume} {53}},\ \bibinfo {pages} {3912}
  (\bibinfo {year} {1996})}\BibitemShut {NoStop}%
\bibitem [{\citenamefont {Moriya}\ \emph {et~al.}(2014)\citenamefont {Moriya},
  \citenamefont {Sawano}, \citenamefont {Hoshi}, \citenamefont {Masubuchi},
  \citenamefont {Shiraki}, \citenamefont {Wild}, \citenamefont {Neumann},
  \citenamefont {Abstreiter}, \citenamefont {Bougeard}, \citenamefont {Koga},\
  and\ \citenamefont {Machida}}]{moriya2014cubic}%
  \BibitemOpen
  \bibfield  {author} {\bibinfo {author} {\bibfnamefont {R.}~\bibnamefont
  {Moriya}}, \bibinfo {author} {\bibfnamefont {K.}~\bibnamefont {Sawano}},
  \bibinfo {author} {\bibfnamefont {Y.}~\bibnamefont {Hoshi}}, \bibinfo
  {author} {\bibfnamefont {S.}~\bibnamefont {Masubuchi}}, \bibinfo {author}
  {\bibfnamefont {Y.}~\bibnamefont {Shiraki}}, \bibinfo {author} {\bibfnamefont
  {A.}~\bibnamefont {Wild}}, \bibinfo {author} {\bibfnamefont {C.}~\bibnamefont
  {Neumann}}, \bibinfo {author} {\bibfnamefont {G.}~\bibnamefont {Abstreiter}},
  \bibinfo {author} {\bibfnamefont {D.}~\bibnamefont {Bougeard}}, \bibinfo
  {author} {\bibfnamefont {T.}~\bibnamefont {Koga}}, \ and\ \bibinfo {author}
  {\bibfnamefont {T.}~\bibnamefont {Machida}},\ }\href@noop {} {\bibfield
  {journal} {\bibinfo  {journal} {Physical review letters}\ }\textbf {\bibinfo
  {volume} {113}},\ \bibinfo {pages} {086601} (\bibinfo {year}
  {2014})}\BibitemShut {NoStop}%
\bibitem [{\citenamefont {Winkler}\ \emph {et~al.}(2008)\citenamefont
  {Winkler}, \citenamefont {Culcer}, \citenamefont {Papadakis}, \citenamefont
  {Habib},\ and\ \citenamefont {Shayegan}}]{0268-1242-23-11-114017}%
  \BibitemOpen
  \bibfield  {author} {\bibinfo {author} {\bibfnamefont {R.}~\bibnamefont
  {Winkler}}, \bibinfo {author} {\bibfnamefont {D.}~\bibnamefont {Culcer}},
  \bibinfo {author} {\bibfnamefont {S.~J.}\ \bibnamefont {Papadakis}}, \bibinfo
  {author} {\bibfnamefont {B.}~\bibnamefont {Habib}}, \ and\ \bibinfo {author}
  {\bibfnamefont {M.}~\bibnamefont {Shayegan}},\ }\href
  {http://stacks.iop.org/0268-1242/23/i=11/a=114017} {\bibfield  {journal}
  {\bibinfo  {journal} {Semiconductor Science and Technology}\ }\textbf
  {\bibinfo {volume} {23}},\ \bibinfo {pages} {114017} (\bibinfo {year}
  {2008})}\BibitemShut {NoStop}%
\bibitem [{\citenamefont {Iordanskii}\ \emph {et~al.}(1994)\citenamefont
  {Iordanskii}, \citenamefont {Lyanda-Geller},\ and\ \citenamefont
  {Pikus}}]{iordanskii1994weak}%
  \BibitemOpen
  \bibfield  {author} {\bibinfo {author} {\bibfnamefont {S.}~\bibnamefont
  {Iordanskii}}, \bibinfo {author} {\bibfnamefont {Y.~B.}\ \bibnamefont
  {Lyanda-Geller}}, \ and\ \bibinfo {author} {\bibfnamefont {G.}~\bibnamefont
  {Pikus}},\ }\href@noop {} {\bibfield  {journal} {\bibinfo  {journal} {ZhETF
  Pisma Redaktsiiu}\ }\textbf {\bibinfo {volume} {60}},\ \bibinfo {pages} {199}
  (\bibinfo {year} {1994})}\BibitemShut {NoStop}%
\bibitem [{\citenamefont {Zudov}\ \emph {et~al.}(2014)\citenamefont {Zudov},
  \citenamefont {Mironov}, \citenamefont {Ebner}, \citenamefont {Martin},
  \citenamefont {Shi},\ and\ \citenamefont {Leadley}}]{PhysRevB.89.125401}%
  \BibitemOpen
  \bibfield  {author} {\bibinfo {author} {\bibfnamefont {M.~A.}\ \bibnamefont
  {Zudov}}, \bibinfo {author} {\bibfnamefont {O.~A.}\ \bibnamefont {Mironov}},
  \bibinfo {author} {\bibfnamefont {Q.~A.}\ \bibnamefont {Ebner}}, \bibinfo
  {author} {\bibfnamefont {P.~D.}\ \bibnamefont {Martin}}, \bibinfo {author}
  {\bibfnamefont {Q.}~\bibnamefont {Shi}}, \ and\ \bibinfo {author}
  {\bibfnamefont {D.~R.}\ \bibnamefont {Leadley}},\ }\href {\doibase
  10.1103/PhysRevB.89.125401} {\bibfield  {journal} {\bibinfo  {journal} {Phys.
  Rev. B}\ }\textbf {\bibinfo {volume} {89}},\ \bibinfo {pages} {125401}
  (\bibinfo {year} {2014})}\BibitemShut {NoStop}%
\bibitem [{\citenamefont {Laroche}\ \emph {et~al.}(2016)\citenamefont
  {Laroche}, \citenamefont {Huang}, \citenamefont {Chuang}, \citenamefont {Li},
  \citenamefont {Liu},\ and\ \citenamefont {Lu}}]{doi:10.1063/1.4953399}%
  \BibitemOpen
  \bibfield  {author} {\bibinfo {author} {\bibfnamefont {D.}~\bibnamefont
  {Laroche}}, \bibinfo {author} {\bibfnamefont {S.-H.}\ \bibnamefont {Huang}},
  \bibinfo {author} {\bibfnamefont {Y.}~\bibnamefont {Chuang}}, \bibinfo
  {author} {\bibfnamefont {J.-Y.}\ \bibnamefont {Li}}, \bibinfo {author}
  {\bibfnamefont {C.~W.}\ \bibnamefont {Liu}}, \ and\ \bibinfo {author}
  {\bibfnamefont {T.~M.}\ \bibnamefont {Lu}},\ }\href {\doibase
  10.1063/1.4953399} {\bibfield  {journal} {\bibinfo  {journal} {Applied
  Physics Letters}\ }\textbf {\bibinfo {volume} {108}},\ \bibinfo {pages}
  {233504} (\bibinfo {year} {2016})},\ \Eprint
  {http://arxiv.org/abs/http://dx.doi.org/10.1063/1.4953399}
  {http://dx.doi.org/10.1063/1.4953399} \BibitemShut {NoStop}%
\bibitem [{\citenamefont {Elliott}(1954)}]{elliott1954theory}%
  \BibitemOpen
  \bibfield  {author} {\bibinfo {author} {\bibfnamefont {R.~J.}\ \bibnamefont
  {Elliott}},\ }\href@noop {} {\bibfield  {journal} {\bibinfo  {journal}
  {Physical Review}\ }\textbf {\bibinfo {volume} {96}},\ \bibinfo {pages} {266}
  (\bibinfo {year} {1954})}\BibitemShut {NoStop}%
\bibitem [{\citenamefont {Yafet}(1983)}]{yafet1983conduction}%
  \BibitemOpen
  \bibfield  {author} {\bibinfo {author} {\bibfnamefont {Y.}~\bibnamefont
  {Yafet}},\ }\href@noop {} {\bibfield  {journal} {\bibinfo  {journal} {Physics
  Letters A}\ }\textbf {\bibinfo {volume} {98}},\ \bibinfo {pages} {287}
  (\bibinfo {year} {1983})}\BibitemShut {NoStop}%
\bibitem [{\citenamefont {Dyakonov}\ and\ \citenamefont
  {Perel}(1972)}]{dyakonov1972spin}%
  \BibitemOpen
  \bibfield  {author} {\bibinfo {author} {\bibfnamefont {M.}~\bibnamefont
  {Dyakonov}}\ and\ \bibinfo {author} {\bibfnamefont {V.}~\bibnamefont
  {Perel}},\ }\href@noop {} {\bibfield  {journal} {\bibinfo  {journal} {Soviet
  Physics Solid State, Ussr}\ }\textbf {\bibinfo {volume} {13}},\ \bibinfo
  {pages} {3023} (\bibinfo {year} {1972})}\BibitemShut {NoStop}%
\bibitem [{\citenamefont {Minkov}\ \emph {et~al.}(2004)\citenamefont {Minkov},
  \citenamefont {Germanenko}, \citenamefont {Rut}, \citenamefont
  {Sherstobitov}, \citenamefont {Golub}, \citenamefont {Zvonkov},\ and\
  \citenamefont {Willander}}]{minkov2004weak}%
  \BibitemOpen
  \bibfield  {author} {\bibinfo {author} {\bibfnamefont {G.}~\bibnamefont
  {Minkov}}, \bibinfo {author} {\bibfnamefont {A.}~\bibnamefont {Germanenko}},
  \bibinfo {author} {\bibfnamefont {O.}~\bibnamefont {Rut}}, \bibinfo {author}
  {\bibfnamefont {A.}~\bibnamefont {Sherstobitov}}, \bibinfo {author}
  {\bibfnamefont {L.}~\bibnamefont {Golub}}, \bibinfo {author} {\bibfnamefont
  {B.}~\bibnamefont {Zvonkov}}, \ and\ \bibinfo {author} {\bibfnamefont
  {M.}~\bibnamefont {Willander}},\ }\href@noop {} {\bibfield  {journal}
  {\bibinfo  {journal} {Physical Review B}\ }\textbf {\bibinfo {volume} {70}},\
  \bibinfo {pages} {155323} (\bibinfo {year} {2004})}\BibitemShut {NoStop}%
\bibitem [{\citenamefont {Fal'ko}(1990)}]{0953-8984-2-16-009}%
  \BibitemOpen
  \bibfield  {author} {\bibinfo {author} {\bibfnamefont {V.~I.}\ \bibnamefont
  {Fal'ko}},\ }\href {http://stacks.iop.org/0953-8984/2/i=16/a=009} {\bibfield
  {journal} {\bibinfo  {journal} {Journal of Physics: Condensed Matter}\
  }\textbf {\bibinfo {volume} {2}},\ \bibinfo {pages} {3797} (\bibinfo {year}
  {1990})}\BibitemShut {NoStop}%
\bibitem [{\citenamefont {Meyer}\ \emph {et~al.}(2002)\citenamefont {Meyer},
  \citenamefont {Altland},\ and\ \citenamefont
  {Altshuler}}]{PhysRevLett.89.206601}%
  \BibitemOpen
  \bibfield  {author} {\bibinfo {author} {\bibfnamefont {J.~S.}\ \bibnamefont
  {Meyer}}, \bibinfo {author} {\bibfnamefont {A.}~\bibnamefont {Altland}}, \
  and\ \bibinfo {author} {\bibfnamefont {B.~L.}\ \bibnamefont {Altshuler}},\
  }\href {\doibase 10.1103/PhysRevLett.89.206601} {\bibfield  {journal}
  {\bibinfo  {journal} {Phys. Rev. Lett.}\ }\textbf {\bibinfo {volume} {89}},\
  \bibinfo {pages} {206601} (\bibinfo {year} {2002})}\BibitemShut {NoStop}%
\bibitem [{\citenamefont {Zumb\"uhl}\ \emph {et~al.}(2002)\citenamefont
  {Zumb\"uhl}, \citenamefont {Miller}, \citenamefont {Marcus}, \citenamefont
  {Campman},\ and\ \citenamefont {Gossard}}]{PhysRevLett.89.276803}%
  \BibitemOpen
  \bibfield  {author} {\bibinfo {author} {\bibfnamefont {D.~M.}\ \bibnamefont
  {Zumb\"uhl}}, \bibinfo {author} {\bibfnamefont {J.~B.}\ \bibnamefont
  {Miller}}, \bibinfo {author} {\bibfnamefont {C.~M.}\ \bibnamefont {Marcus}},
  \bibinfo {author} {\bibfnamefont {K.}~\bibnamefont {Campman}}, \ and\
  \bibinfo {author} {\bibfnamefont {A.~C.}\ \bibnamefont {Gossard}},\ }\href
  {\doibase 10.1103/PhysRevLett.89.276803} {\bibfield  {journal} {\bibinfo
  {journal} {Phys. Rev. Lett.}\ }\textbf {\bibinfo {volume} {89}},\ \bibinfo
  {pages} {276803} (\bibinfo {year} {2002})}\BibitemShut {NoStop}%
\bibitem [{\citenamefont {Zhang}\ \emph {et~al.}(2012)\citenamefont {Zhang},
  \citenamefont {Iwasaki}, \citenamefont {Taoka}, \citenamefont {Takenaka},\
  and\ \citenamefont {Takagi}}]{zhang2012high}%
  \BibitemOpen
  \bibfield  {author} {\bibinfo {author} {\bibfnamefont {R.}~\bibnamefont
  {Zhang}}, \bibinfo {author} {\bibfnamefont {T.}~\bibnamefont {Iwasaki}},
  \bibinfo {author} {\bibfnamefont {N.}~\bibnamefont {Taoka}}, \bibinfo
  {author} {\bibfnamefont {M.}~\bibnamefont {Takenaka}}, \ and\ \bibinfo
  {author} {\bibfnamefont {S.}~\bibnamefont {Takagi}},\ }\href@noop {}
  {\bibfield  {journal} {\bibinfo  {journal} {IEEE Transactions on Electron
  Devices}\ }\textbf {\bibinfo {volume} {59}},\ \bibinfo {pages} {335}
  (\bibinfo {year} {2012})}\BibitemShut {NoStop}%
\end{thebibliography}
\end{document}